\documentclass[amsmath,amssymb,aps,prd,superscriptaddress,
reprint
]{revtex4-1}

\usepackage{empheq}
\usepackage{textcomp}
\usepackage{bigints}
\usepackage{colortbl}
\usepackage{multirow}
\usepackage{array}
\usepackage{graphicx}
\usepackage{braket}
\usepackage[table,xcdraw]{xcolor}
\usepackage[mathlines]{lineno}
\usepackage{float}
\usepackage{bigints}
\usepackage{makecell}
\usepackage{graphicx}
\usepackage{dcolumn}
\usepackage{bm}
\usepackage{hyperref}
\hypersetup{
    colorlinks = true,
    linkcolor = black,
    citecolor=blue, 
    }
\usepackage[capitalise]{cleveref}
\usepackage{siunitx}
\usepackage{graphicx} 
\usepackage{booktabs}
\usepackage{chemformula}
\usepackage{upgreek}
\usepackage[mathlines]{lineno}
\begin{document}

\title{Fourth-harmonic UV light generation in integrated silicon nitride microresonators}

\author{Alekhya Ghosh}
\thanks{These authors contributed equally}
\affiliation{Max Planck Institute for the Science of Light, D-91058 Erlangen, Germany}
\affiliation{Department of Physics, Friedrich Alexander University Erlangen-Nuremberg, D-91058 Erlangen, Germany}

\author{Arghadeep Pal}
\thanks{These authors contributed equally}
\affiliation{Max Planck Institute for the Science of Light, D-91058 Erlangen, Germany}
\affiliation{Department of Physics, Friedrich Alexander University Erlangen-Nuremberg, D-91058 Erlangen, Germany}

\author{Haochen Yan}
\affiliation{Max Planck Institute for the Science of Light, D-91058 Erlangen, Germany}
\affiliation{Department of Physics, Friedrich Alexander University Erlangen-Nuremberg, D-91058 Erlangen, Germany}

\author{Toby Bi}
\affiliation{Max Planck Institute for the Science of Light, D-91058 Erlangen, Germany}
\affiliation{Department of Physics, Friedrich Alexander University Erlangen-Nuremberg, D-91058 Erlangen, Germany}

\author{Luca O. Trinchão}
\affiliation{Max Planck Institute for the Science of Light, D-91058 Erlangen, Germany}
\affiliation{Gleb Wataghin Institute of Physics, University of Campinas, Campinas, SP, Brazil}

\author{Qixuan Zhou}
\affiliation{Max Planck Institute for the Science of Light, D-91058 Erlangen, Germany}
\affiliation{Department of Physics, Friedrich Alexander University Erlangen-Nuremberg, D-91058 Erlangen, Germany}

\author{Gustavo S. Wiederhecker}
\affiliation{Gleb Wataghin Institute of Physics, University of Campinas, Campinas, SP, Brazil}

\author{Shuangyou Zhang}
\affiliation{Department of Electrical and Photonics Engineering Technical University of Denmark Kgs. Lyngby 2800, Denmark}

\author{Pascal Del’Haye}
\email[e-mail:]{ pascal.delhaye@mpl.mpg.de}
\affiliation{Max Planck Institute for the Science of Light, D-91058 Erlangen, Germany}
\affiliation{Department of Physics, Friedrich Alexander University Erlangen-Nuremberg, D-91058 Erlangen, Germany}

\date{\today}

\begin{abstract}

Integrated silicon nitride (Si$_3$N$_4$) resonators have emerged as a leading platform for nonlinear photonics, yet generating light at wavelength in the ultraviolet (UV) has remained elusive in single-resonator systems.
Here we report the first observation of fourth-harmonic generation reaching the blue and ultraviolet spectral regions in an integrated Si$_3$N$_4$ microring resonator. 
We systematically investigate the input-power dependence of the wavelength ranges supporting second-, third-, and fourth-harmonic generation, and study the input-power-dependent variation of the circulating fourth-harmonic signal in the UV.
These results extend the operational bandwidth of integrated Si$_3$N$_4$ nonlinear photonic platforms to the lower edge of the material transparency window, enabling on-chip UV frequency conversion. Near-ultraviolet generation around $400$~nm will enable on-chip excitation of defect-based quantum emitters in hexagonal boron nitride, enhance Raman spectroscopy through increased scattering cross-sections at shorter wavelengths, and support compact fluorescence-based bio-imaging platforms exploiting intrinsic cellular fluorophores.
\end{abstract}

\maketitle

\section{Introduction}

\indent Generating light with frequencies in the visible and ultra-violet (UV) ranges in integrated platforms are essential for on-chip optical clocks~\cite{ludlow2015optical}, metrology, telecommunication~\cite{pathak2015visible, tua2023imaging}, spectroscopy~\cite{tua2023imaging}, quantum information processing~\cite{mehta2020integrated}, biomedical imaging systems~\cite{sacher2021implantable}, laser displays~\cite{zhao2019full}, and quantum sensing~\cite{lawrie2019quantum}. 
Ultra-low propagation loss, CMOS-compatible fabrication, and a broad transparency window have established silicon nitride (Si$_3$N$_4$) as one of the leading platforms for integrated photonics~\cite{xiang2022silicon}. Ongoing advances in fabrication have produced Si$_3$N$_4$ ring resonators with exceptionally high quality factors~\cite{zhang2024low, puckett2021422}. 
These improvements have enabled generation of temporal cavity solitons~\cite{ji2025deterministic}, and spontaneous symmetry breaking~\cite{zhang2025integrated, trinchao2026color} in Si$_3$N$_4$ resonators. Together, these breakthroughs are driving applications ranging from LiDAR~\cite{trocha2018ultrafast} and precision spectroscopy~\cite{dutt2018chip} to high-capacity telecommunications~\cite{pfeifle2014coherent}.

\begin{figure*}[t]
\centering
\includegraphics[width=\textwidth]{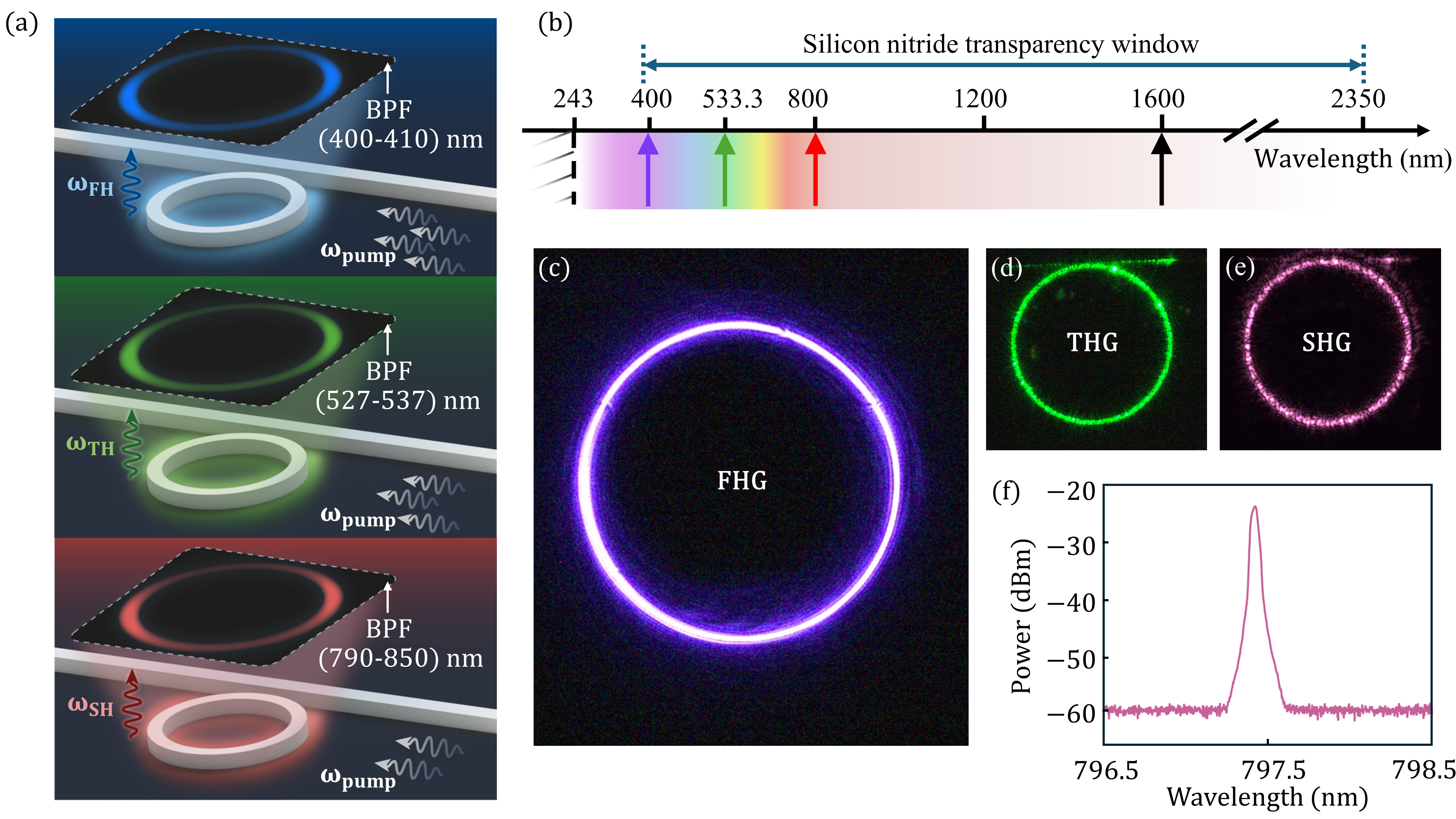}
\caption{\label{Fig:1}
{\textit{Harmonic (second, third and fourth) generation in a single microresonator.} (a) Schematics showing the fourth- (top), third- (middle) and second- (bottom) harmonic light generated in the cavity. The harmonic light is measured with a camera after applying bandpass filters (BPF) with transmission windows indicated in the figure. (b) With a pump wavelength of 1595 nm, we can reach the lower wavelength edge of the silicon nitride (Si$_3$N$_4$) transparency window by generating a fourth-harmonic signal. Although the intrinsic electronic bandgap of Si$_3$N$_4$ corresponds to a cut-off wavelength of approximately 243 nm ($\approx$5 eV), sub-bandgap absorption arising from band-tail states and defect-related electronic states shifts the practical optical transparency edge to around 400 nm in deposited films. (c) Image of the fourth-harmonic generation (FHG) measured with a camera with a bandpass filter at 405 nm with 10 nm bandwidth. (d) Third-harmonic generation (THG) captured through a bandpass filter at 532 nm with 10 nm bandwidth. (e) Second-harmonic image using a bandpass filter with a pass band from 790 nm to 850 nm. (f) Optical spectrum analyzer trace of the second harmonic signal.
}
}
\end{figure*}
Integrating laser sources or active media on Si$_3$N$_4$ chips enables generations of light in the UV~\cite{franken2023hybrid}, visible~\cite{corato2023widely, winkler2024widely} and infra-red~\cite{yang2024titanium} spectral ranges, however, this require complex fabrication processes. A possible method to access new frequencies in integrated Si$_3$N$_4$ platforms is to exploit Kerr-nonlinear frequency conversion~\cite{lu2022kerr, stone2024chip, li2016efficient, sun2024advancing, perez2023high, raghunathan2025telecom, pal2025hybrid}. 
Recent research also addressed accessing visible wavelengths by Kerr-interactions~\cite{corato2025simultaneous}, second-harmonic generation (SHG) and third-harmonic generation (THG)~\cite{yuan2025efficient, levy2011harmonic, lu2021efficient, li2023high, franken2503milliwatt, sayem2021efficient}. 
\begin{figure*}[ht!]
\centering
\includegraphics[width=2\columnwidth]{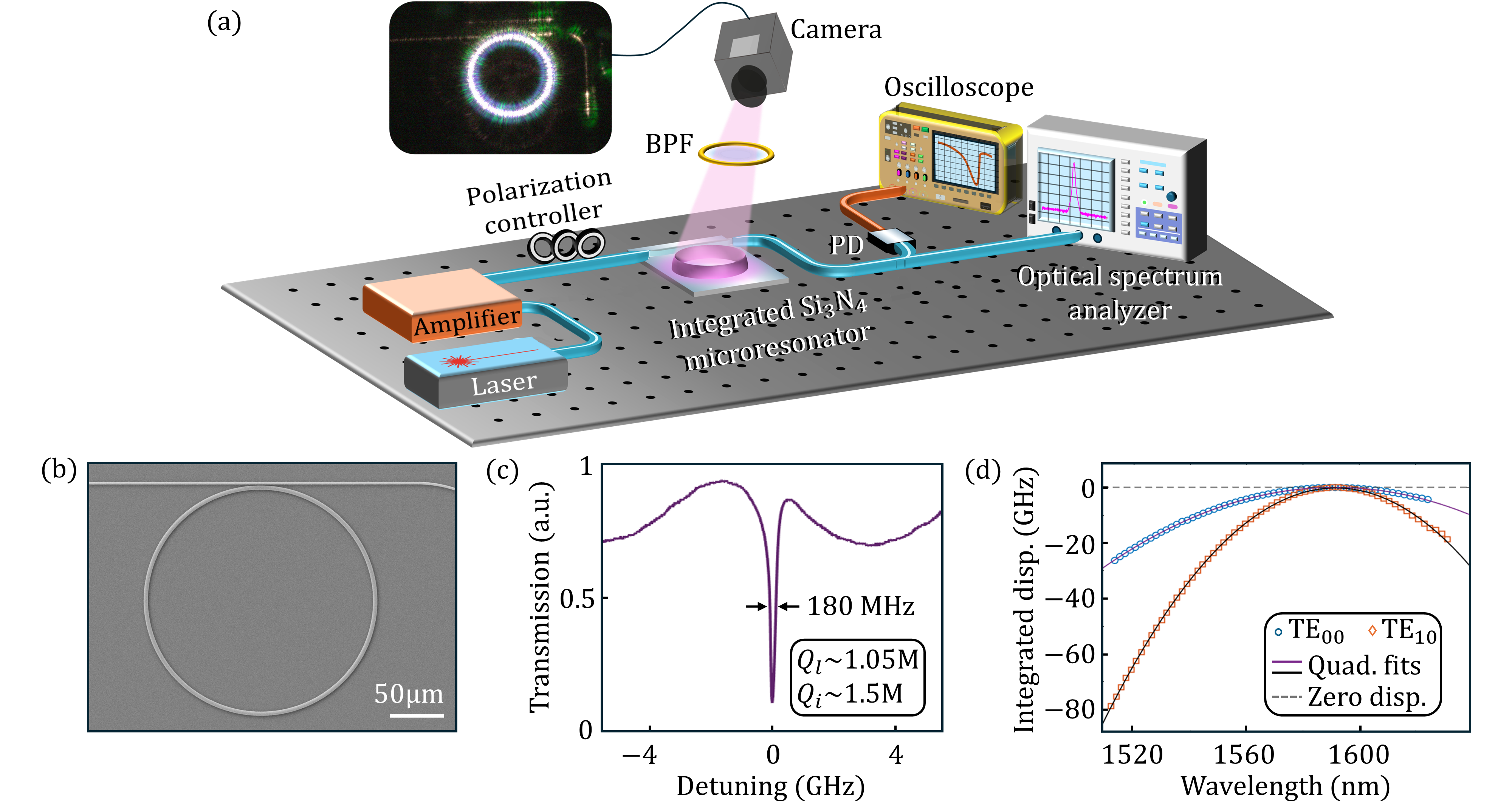}
\caption{\textit{Experimental setup and resonator characterization.} (a) Experimental setup with an inset camera image that shows the generation of all three harmonics simultaneously in a single Si$_3$N$_4$ microring resonator. PD: Photodiode, BPF: Bandpass filter. (b) Scanning electron microscope image of the Si$_3$N$_4$ ring resonator. (c) Transmission profile of a resonance at $1591$~nm showing its linewidth and quality factor. (d) Measured integrated dispersion as a function of wavelength for both the fundamental and higher-order spatial modes. The solid lines show the quadratic fits to the dispersion data points for each mode along with the zero dispersion line shown as gray dashes.}
\label{Fig_2_manuscript}
\end{figure*}
In addition, a two-dimensional $10 \times 10$ microresonator array has demonstrated fourth-harmonic (FH) blue-light generation under high peak power pulsed pumping conditions~\cite{mehrabad2025multi}.\\
\indent In this work, we demonstrate, to the best of our knowledge, for the first time, the generation of UV light via fourth-harmonic generation (FHG) in a high-Q integrated Si$_3$N$_4$ resonator under continuous wave pumping. Figure~\ref{Fig:1}(a) shows a schematic picture showing the generation of second-harmonic (SH), third-harmonic (TH) as well as FH signals in a single resonator observed from a camera after the light passes through suitable filters. A normal dispersion resonator is pumped in the L-band and at different detunings SH, TH and FH are generated. By adjusting the input power and pump frequency, various combinations of harmonic generation can be realized, including the simultaneous generation of all three harmonics. Our work pushes the nonlinear optical wavelength conversion to the UV at the edge of the transparency range of Si$_3$N$_4$ (Fig.~\ref{Fig:1}(b)). Figure~\ref{Fig:1}(c), (d) and (e) display the scattered FH UV light, TH green light and SH infrared light captured with a camera. The spectra of the corresponding SH light is shown in Fig.~\ref{Fig:1}(f). The results from our work can be useful for 
label-free biomedical imaging~\cite{sparks2015flexible}, optical atomic clocks~\cite{king2022optical}, light generation for optical projectors, high-resolution imaging with entangled photon pairs~\cite{edamatsu2004generation}, and portable chemical sensors~\cite{liu2019beyond}. In integrated photonics, blue and UV light can further facilitate tighter waveguide bends.
\section{Results}
\subsection{Experimental setup}
\begin{figure*}[ht!]
\centering
\includegraphics[width=\linewidth]{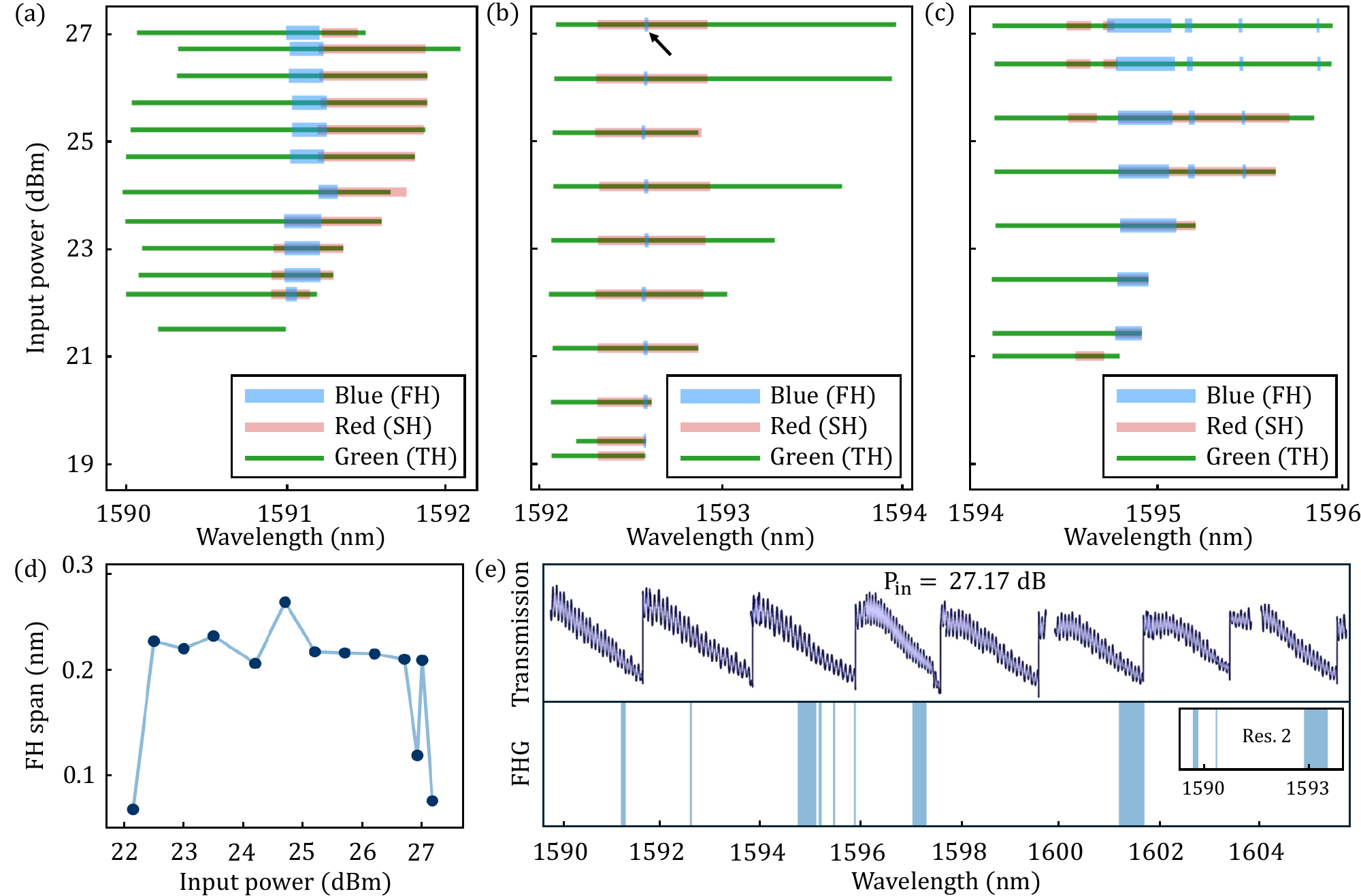}
\caption{\textit{Phase-matching regions for harmonic generation.} (a-c) Wavelength ranges supporting generation of second, third, and fourth-harmonics for different on-chip input powers. The corresponding cold-cavity resonances are at (a) 1590.05~nm, (b) 1592.07~nm and (c) 1594.09~nm. The black arrow in panel (b) indicates the region supporting fourth-harmonic generation during the detuning scan at the highest input power; it is included for clarity. (d) Variation of the detuning span of the fourth-harmonic light at different input powers for a resonance near 1590 nm. (e) Transmission spectrum of several consecutive resonator modes (top row) and corresponding wavelength ranges supporting FHG (bottom row). The inset in the bottom row shows the reproducibility of FHG in another resonator of similar dimensions, despite fabrication-induced variations. FH: fourth-harmonic; TH: third-harmonic; SH: second-harmonic.}
\label{Fig:3}
\end{figure*}
Figure~\ref{Fig_2_manuscript}(a) shows the experimental setup, where a tunable continuous wave laser (ECDL) is amplified with an L-band Erbium doped fiber amplifier (EDFA) and coupled to a chip with an integrated Si$_3$N$_4$ resonator with a lensed fiber. 
The output from the resonator is coupled back into a waveguide and then collected with another lensed fiber. 
A part of the output light is sent to an optical spectrum analyzer (OSA) for spectral analysis, while the rest is sent to a photodetector (PD) that is monitored with an oscilloscope (OSC).\\
\indent On the chip, a coupling waveguide with cross-section $1~\upmu$m (width)$\times400$~nm (height) couples the light to a ring resonator with a radius of $100~\upmu$m and core cross-section $1.8~\upmu$m (width)$\times400$~nm (height). 
Both the waveguide and the resonator are embedded in silica. The coupling gap between the waveguide and the ring is $500$~nm. 
The resonator is fabricated using the process detailed in Ref.~\cite{zhang2024low}. 
Figure~\ref{Fig_2_manuscript}(b) shows a scanning electron microscope image of the resonator before cladding deposition. 
The resonator has a free spectral range of $239~$GHz and the intrinsic quality factor is estimated to be $1.5$~million from the transmission profile of the fundamental TE$_{00}$ mode of the resonator at 1591 nm, shown in Fig.~\ref{Fig_2_manuscript}(c). 
The dispersion of the resonator is normal for both the fundamental mode as well as the higher order TE$_{10}$ mode as shown in Fig.~\ref{Fig_2_manuscript}(d).\\
\indent Since the coupling waveguide is designed to efficiently couple light around $1550$~nm, the coupling efficiency gradually decreases with decreasing wavelength, reducing significantly near the SH, and even more near the TH and FH signals.
When the resonator is pumped at 1595 nm, simulations show that the out-coupling coefficients from the resonator to the bus waveguide for the SH, TH, and FH signals are 4.74~\% (-13.24~dB), 0.0003~\%(-55.23~dB), and 0.00005~\%(-63.01~dB), respectively, relative to the coupling at the pump wavelength.
Thus, the TH and FH signals are monitored via the upward-scattered light from the resonator.
A camera is positioned vertically above the resonator with an imaging system consisting of a magnification tube (6.5~$\times$ with 0.006 to 0.142 system NA) and a magnification lens (2.0~$\times$ magnification) for capturing an image of the resonator~\cite{yan2024real, zhang2025visualization}. 
Bandpass filters (BPFs) are inserted in between the resonator and imaging system to monitor selected wavelength ranges (corresponding to the harmonic signals). 
For monitoring SH, TH and FH, BPFs with pass-band ranges of 790–850 nm, 527-537~nm, and 400-410~nm are selected, respectively.

\subsection{Fourth-harmonic generation in a Si$_3$N$_4$ resonator}
In the wavelength range from $1589$~nm to $1596$~nm the resonator under study exhibits three TE$_{00}$ resonances. 
At an input power level of $27.17$~dB, we tune the laser to each resonance and record the range of detunings that support red, green and blue light generation through SHG, THG and FHG. We record the detuning ranges at different input power levels. 
The input power is gradually reduced until no significant FHG is observed. 
We show the result in Fig.~\ref{Fig:3}(a), where different colors indicate the detuning ranges that support different harmonics.\\
\indent Three points should be noted at this stage. First, even though the width of the thermal triangles (triangular transmission profiles formed due to temperature-
induced resonance shifts) increases with increasing input power, the detuning ranges that support generation of different harmonics do not trivially follow this trend. 
Increasing input power changes position of resonances near the harmonics by means of thermal effects and cross-phase modulation.
This alters the phase-matching conditions contributing to harmonic generations, as shown in panels (a-c) of Fig.~\ref{Fig:3}. 
Second, the three harmonics can occur simultaneously, individually, or in pairs of any two of them. 
Different phase-matching conditions get fulfilled for different combinations of input parameters, which are difficult to predict or model without prior knowledge of the resonator's spectral characteristics around the harmonic frequencies. 
In distinct wavelength regions of the input laser, FHG is supported without observation of SH, thus indicating that in these regions, the FHG is not generated by cascaded SHG. Third, in a single resonance, there can be multiple isolated sections of detunings that supports different harmonic generations. This can be seen in wavelength scans at some input powers ($>$24 dBm) in Fig.~\ref{Fig:3}(c).
\begin{figure}[t!]
\centering
\includegraphics[width=1\columnwidth]{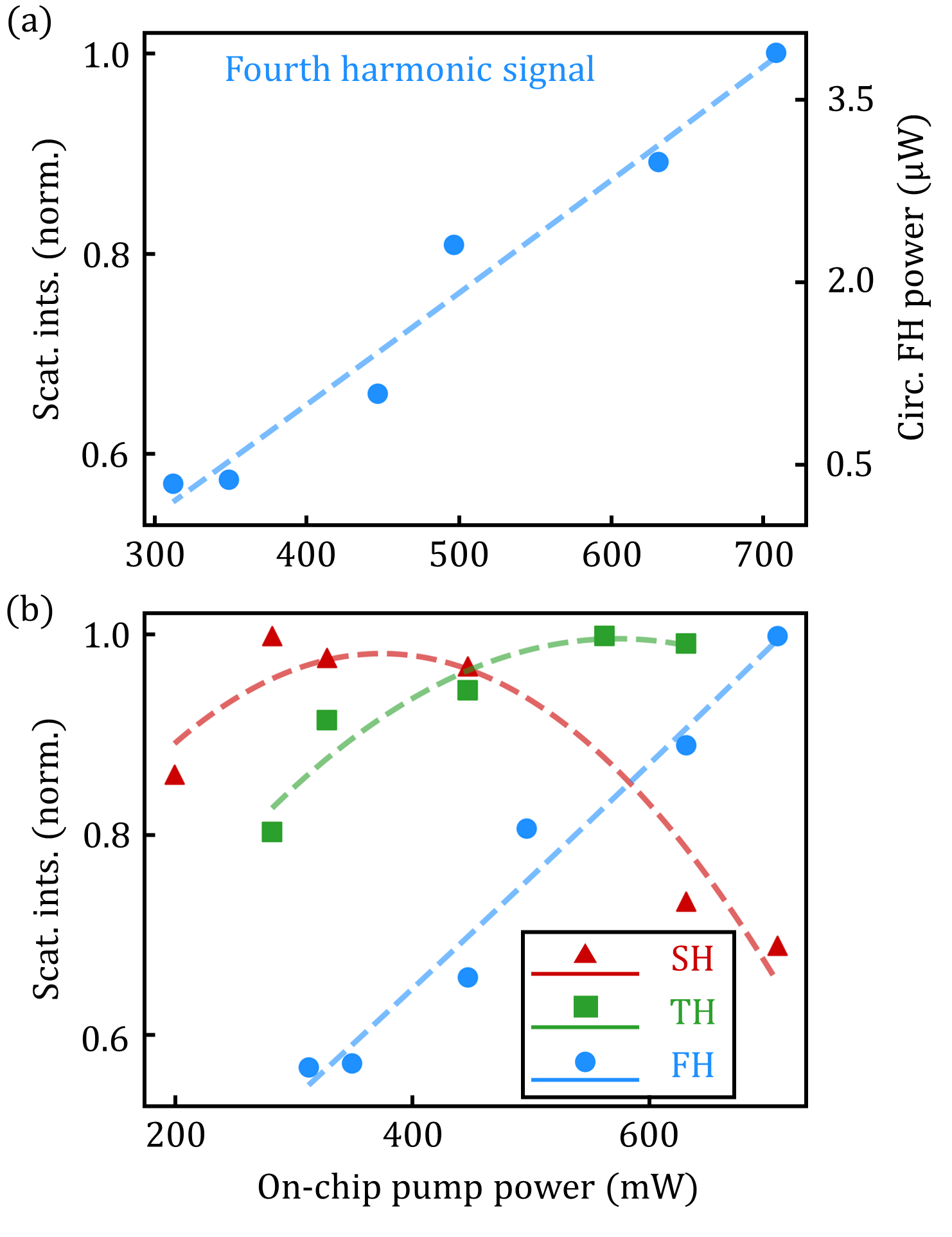}
\caption{\textit{Dependence of harmonic generation on input power.} (a) Normalized scattered intensity of the fourth-harmonic signal as a function of input power, showing a gradual increase. The corresponding circulating power inside the resonator is indicated on the right axis. Panel (b) shows variation of the normalized scattered light intensities of different harmonics with increasing input power. }
\label{Fig:4}
\end{figure}

\indent Figure~\ref{Fig:3}(d) shows the variation of the input laser's wavelength range that supports FHG as a function of input power. It can be observed that the wavelength span supporting FHG varies nonlinear with the input power, which is attributed to the different phase-matching conditions.\\
\indent The transmission profile of the Si$_3$N$_4$ resonator, depicted in the top panel of Fig.~\ref{Fig:3}(e) shows a scan across 8 consecutive microresonator resonances at an input power (on-chip) of 27.17~dB. The lower panel shows the wavelength ranges supporting FHG. The inset shows the wavelength range for a second resonator with identical design dimensions. 
The difference in the frequency ranges supporting FHG originates from the fabrication induced tolerances resulting in different phase-matching conditions.

\indent Figure~\ref{Fig:4}(a) shows that the normalized scattered intensity (and the corresponding estimated circulating power) of the FH signal increases with increasing input pump power. The method to estimate the circulating FH power is described in the supplementary material. 
The scattered light intensities of all generated harmonics as a function of the on-chip input power are shown in Fig.~\ref{Fig:4}(b). 
When the input power increases, the SH signal decreases while the FH signal increases.
This could indicate an energy transfer from the SH to the FH signal.
We pump the resonator at a cold-cavity wavelength of 1594.07~nm. When the pump power increases, the resonance shifts due to thermal and Kerr effects. Thus, the detuning is adjusted to maximize the FHG process, and the data for all harmonics are recorded at the detunings corresponding to maximum FHG.
In contrast, the THG efficiency saturates at higher power levels. 
These trends suggest that the observed FHG may originate from multiple nonlinear pathways, including higher-order $\chi^{(4)}$ processes, cascaded $\chi^{(2)}$ interactions, and combined $\chi^{(2)}$ and $\chi^{(3)}$ nonlinear mechanisms. 

\subsection{Phase-matching and spatial mode overlap}
\begin{figure*}[ht]
\centering
\includegraphics[width=1\linewidth]{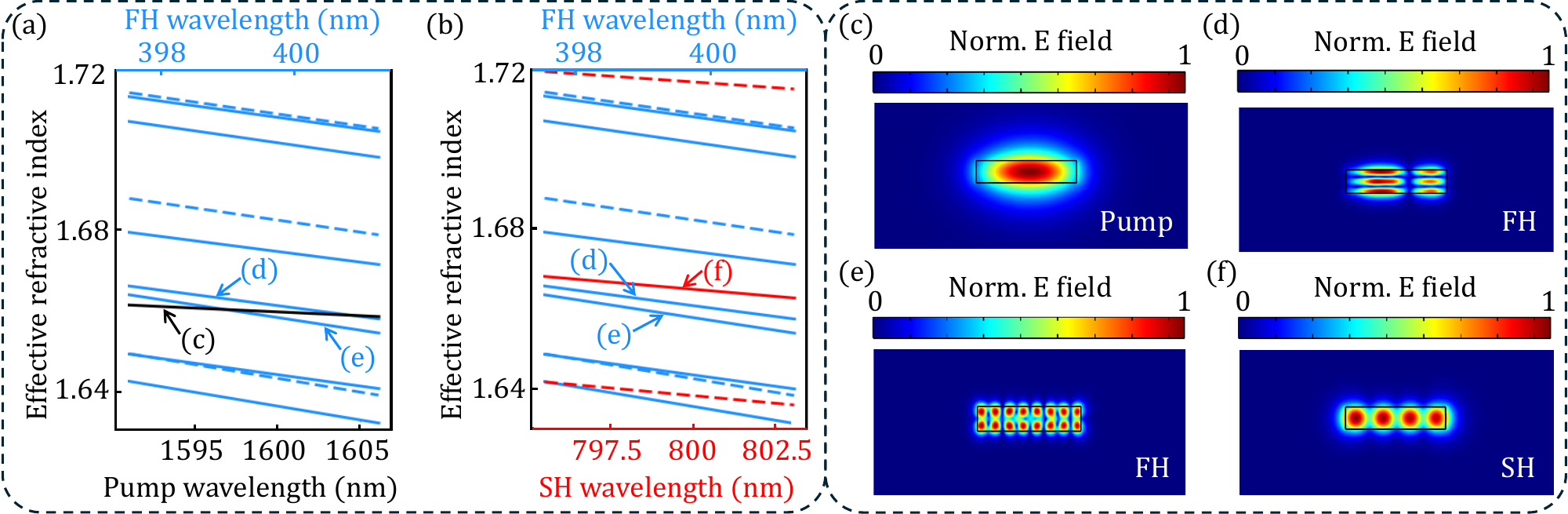}
\caption{\textit{Phase-matching conditions for harmonic generation.} Panel (a) shows effective refractive indices as a function of wavelength showing phase-matching between the available TE modes at 1550 nm (in black color) and higher-order modes at fourth-harmonic (FH).
Panel (b) depicts effective index matching between SH and FH modes. Solid and dashed curves denote TE and TM modes, respectively, with colors indicating harmonic orders (blue: FH, red: SH). Panels (c-e) show mode profiles of the pump TE$_{00}$ mode (c) and the corresponding phase-matched modes (effective refractive index of the harmonics close to that at the pump) at (d-e) FH, and (f) SH wavelengths.}
\label{simulation_fig}
\end{figure*}
Efficient generation of $n^\text{th}$ harmonic requires phase-matching, i.e., matching of effective refractive indices at the pump wavelength and the $n^\text{th}$ harmonic wavelength~\cite{soares2023third, rehan2025second}. 
In the experiments, we pump the TE$_{00}$ mode of the resonator.
Figures~\ref{simulation_fig}(a) shows the effective refractive indices of the fundamental pump mode (shown by a black line) in wavelength range from 1590~nm to 1605~nm together with higher-order modes at  FH wavelengths.
The effective refractive indices of the different spatial modes within the corresponding wavelength ranges of SH and FH are shown by red, and blue lines in panel (a-b), respectively. 
It can be seen that the effective refractive indices of the TE$_{00}$ modes are close to the effective refractive indices of various spatial modes for different harmonic signals.
The mode with the closest effective refractive index to that of the pump mode facilitates the corresponding harmonic generation.
However, it should be noted that these plots corresponds to cold-cavity conditions. 
With increasing input power, the thermal and Kerr nonlinearities induce changes of the effective refractive indices, which can significantly alter and often relaxes the phase-matching conditions~\cite{lu2021efficient}.
Moreover, changing the detuning of the input laser can lead to sequential phase-matching with different higher order spatial modes of the harmonic signal, as shown in wavelength scans $>$24dBm in Fig.~\ref{Fig:3}(c). 
Figure~\ref{simulation_fig}(b) shows a comparison of the effective refractive indices of the different spatial modes in the SH and FH wavelength ranges, indicating that in some cases cascaded SHG can lead to the blue light generation.\\
\indent Figure~\ref{simulation_fig}(c) shows mode profile of the pumped TE$_{00}$ mode, while panels (d-e) depict the same for the modes that facilitate the phase-matching for FHG, and SHG respectively.
The phase-matching conditions and the corresponding mode profiles for SHG and THG are shown in the supplementary material. 
The number of available higher order spatial modes increases with shorter wavelength.   
Thus, the chances of phase-matching also increase for higher order harmonics.
However, another important factor contributing to the blue light generation is the spatial mode overlap between the modes involved in nonlinear interactions. The overlap between the pump mode and the higher-harmonic modes decrease with increasing harmonic order, resulting in decreasing efficiencies of light conversion~\cite{rehan2025second}.
Comparing the images of the scattered light, the generated FH is estimated to be around 24~dBm weaker than the SH signal in our resonators.
Further dispersion engineering~\cite{pal2023machine, kheyri2025chip, tomazio2024tunable} of the resonators can enable improved phase-matching between modes with higher spatial overlap.
\section{Discussion and outlook}
We demonstrate, for the first time, fourth-harmonic–induced blue light generation in a single ring resonator on an integrated Si$_3$N$_4$ platform.
Under continuous-wave pumping of the resonator, second-, third-, and fourth-harmonic light are generated.
By studying the wavelength span of the generated harmonics as a function of pump power, we identify doubly-resonant (pump and one harmonic), triply-resonant (pump and two harmonics), and quadruply-resonant (pump and three harmonics) regions.
These observations indicate that fourth-harmonic generation can arise from a superposition of multiple nonlinear processes, including direct $\chi^{(4)}$ interactions, cascaded $\chi^{(2)}+\chi^{(3)}$ processes, and purely cascaded $\chi^{(2)}$ interactions.
The reproducibility of fourth-harmonic generation is observed in two resonators of similar dimensions, although variations in phase-matching conditions arise from fabrication-induced tolerances.
The generation of fourth-harmonic light can bridge the gap between the telecom and visible wavelength ranges, extending down to the UV at the short-wavelength edge of the Si$_3$N$_4$ transparency window, and thereby offering potential applications in sensing, optical clocks, spectroscopy, and quantum optics.
Beyond this demonstration, our future work will investigate materials transparent at shorter wavelengths than Si$_3$N$_4$, such as aluminium nitride~\cite{yan2025simplified} and dispersion engineering for generation of even higher-order harmonic.  
\section*{Data Availability}
Data underlying the results presented in this paper are not publicly available at this time but may be obtained from the authors upon reasonable request.

\section*{Acknowledgements}
AG and AP contributed equally to this work. The authors acknowledge the Chekhova Research Group for lending the bandpass filters.\\
This work was funded by the generation MQV Project TeQSiC, the German Federal Ministry of Research, Technology and Space, Quantum Systems, 13N17314, 13N17342, the Max Planck Society, and the Max Planck School of Photonics. This work was supported by São Paulo Research Foundation (FAPESP) through grants 
18/15580-6, 
18/25339-4, 
21/10334-0, 
23/09412-1, 
24/15935-0, 
25/04049-1, 
25/10683-5, 
and Coordenação de Aperfeiçoamento de Pessoal de Nível Superior - Brasil (CAPES) (Finance Code 001).
SZ and PD acknowledge support from Deutsche Forschungsgemeinschaft project 541267874.

\bibliography{bib.bib}

\onecolumngrid  

\clearpage
\appendix

\section*{Supporting Information}

\normalsize

\setcounter{figure}{0}
\setcounter{equation}{0}
\setcounter{section}{0}

\renewcommand{\thesection}{S\arabic{section}}
\renewcommand{\thefigure}{S\arabic{figure}}
\renewcommand{\theequation}{S\arabic{equation}}

\section{Estimation of fourth-harmonic power based on the measured second-harmonic data}
\label{A}

\begin{figure}[htbp]
    \centering
    \includegraphics[width=\columnwidth]{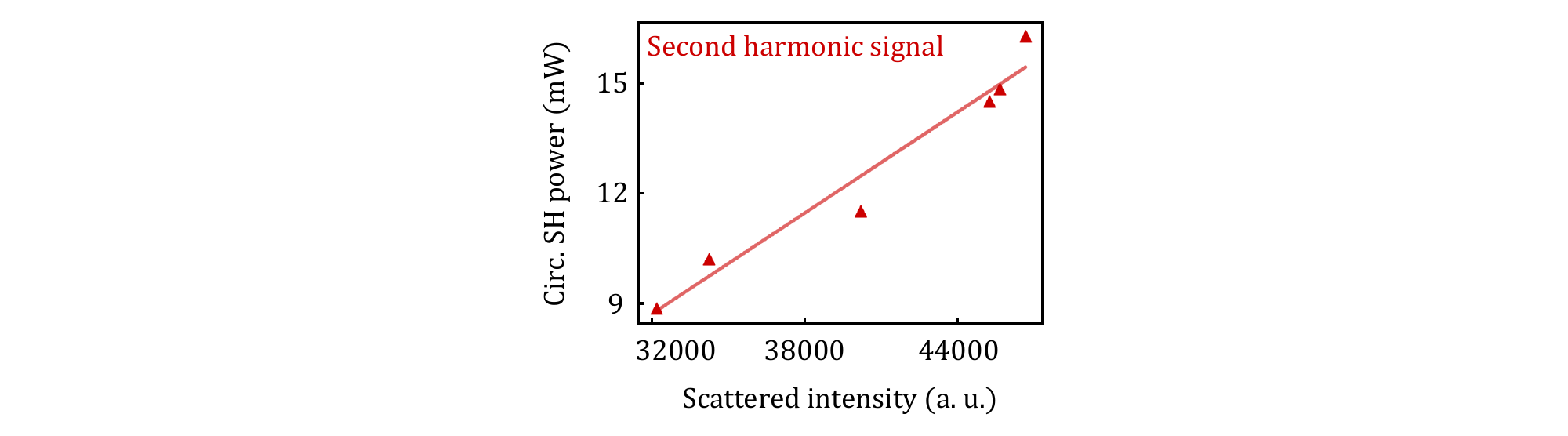}
    \caption{Dependence of circulating power of the second-harmonic signal within
    the resonator as a function of scattered intensity, measured at a pump mode
    with a cold-cavity resonance wavelength of 1594.07~nm.}
    \label{SH_power_vs_Scatt}
\end{figure}

The generated second-harmonic (SH) signal is first measured using an optical spectrum analyzer (OSA). Simultaneously, the upscattered light is monitored with a bandpass filter (BPF) and a camera. The total intensity recorded by the camera is normalized to the camera gain at the SH wavelength. The circulating SH power inside the resonator is estimated by simulating the out-coupling efficiency from the resonator to the bus waveguide. Figure~\ref{SH_power_vs_Scatt} shows a linear relationship between the circulating SH power and the normalized scattered intensity. The slope of the fit, $m_0$, is used as the conversion factor to estimate circulating power from a unit of normalized scattered intensity.\\
\indent The scattered fourth-harmonic (FH) signal is then recorded using a $405\pm5$~nm BPF and a camera. As in the SH case, the captured intensity is normalized to the camera gain. Assuming that the scattered light originates from Rayleigh scattering, for which the scattered intensity scales as $\left(1/\lambda^4\right)$, the normalized scattered intensity corresponding to a unit circulating FH power inside the resonator is considered to be 16 times larger than that of the SH signal. The scattered intensity of SH signal is converted to circulating power using the same scaling factor $m_0$.

\clearpage

\section{Phase-matching at second and third harmonics}
\label{B}

\begin{figure}[htbp]
    \centering
    \includegraphics[width=0.8\columnwidth]{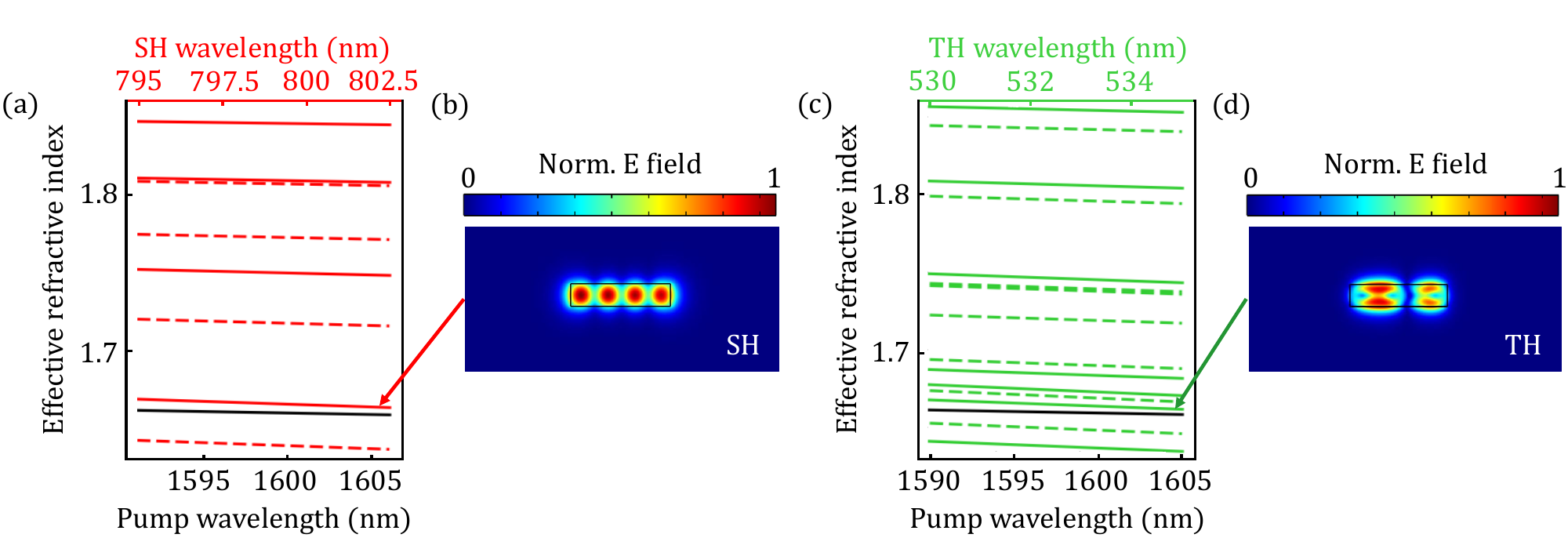}
    \caption{Phase-matching between the pump TE$_{00}$ mode and higher-order mode at (a) second-harmonic (red) wavelength and (c) third-harmonic (green) wavelengths. The solid lines denote transverse electric mode whereas the dashed lines denote the transverse magnetic modes. Panels (b) and (d) show the corresponding spatial mode profiles of SH and TH modes facilitating phase matching.}
    \label{SH_TH}
\end{figure}

Figure~\ref{SH_TH}(a) and (c) depict the overlap between the effective refractive indices of the fundamental mode at the pump wavelengths and those of the higher-order modes at the second- and third-harmonic wavelengths, respectively. 
The effective refractive indices of the fundamental pump modes in the wavelength range of 1590--1605~nm are shown in black. 
The effective refractive indices near the second-harmonic wavelength are shown in red, whereas those near the third-harmonic wavelength are shown in green. 
Solid lines denote transverse electric modes, while dashed lines indicate transverse magnetic modes. At low power, the effective refractive indices of the pump mode lie close to those of the higher-order modes at the harmonic wavelengths. As the power increases, the resonances shift due to thermal and Kerr effects, thereby enhancing the chances of phase-matching and enabling the generation of harmonic signals.
The spatial profile of the SH and TH modes whose effective refractive indices near the respective wavelengths lie close to those of the fundamental mode at the pump wavelength, are shown in Fig.~\ref{SH_TH}(b) and Fig.~\ref{SH_TH}(d) respectively (as indicated by red and green arrows respectively).

\end{document}